# Exchange-Driven Magnetic Excitation and Integrated Magnetoelectronics

1/3/00

J. C. Slonczewski

IBM Research Center

P. O. Box 218, Yorktown Heights, NY 10598

USA

Theory and recent experiments concerning exchange-driven magnetic excitation (EDME) are reviewed.  This phenomenon employs the exchange field produced by a narrowly distributed spin-polarized electron current to excite Larmor precession in a magnetic film or particle.  Predicted threshold currents for such excitation of both two-dimensional spin-waves and of monodomain reversal are now experimentally supported at both helium and ambient temperatures.  The present status of this field suggests a high potential for applications of EDME to the write operation of magnetic recording and, when combined with tunneling magnetoresistance, to a memory latch using sub-200 nm lithography.  This potential is buoyed by very recent experiments at Cornell University implying the theoretical availability of almost hundred-fold advantage in excitation efficiency favoring exchange fields over Maxwell fields.

**Introduction.** A ten-year outlook on future information technology projects pervasive computing embodied in billions of specialized (post-PC) pocketable controller-communicators serving all sorts of appliances and personal needs. The practical need to integrate any potential magnetic function like memory with silicon logic and support circuitry will favor novel magnetic elements that function at ambient temperature, measure < 200 nm, switch in less than 10 ns, draw currents of less than 10 mA, and create signals exceeding 50 mV.

Two recent complementary breakthroughs in the physics of magnetism provide fresh research opportunities in this arena. One is tunneling magnetoresistance (TMR) and the other is exchange-driven (previously called "current-driven" or "spin-transfer") magnetic excitation (EDME). The first breakthrough came slowly; in the beginning TMR was observed only at helium temperatures [1]. An early device concept was a sensor for bubble memory [2]. But nineteen more years passed before substantial TMR effects were observed at room temperature [3]. Subsequent work has produced magnetic resistance changes and signal voltages competing very well with those of giant magneto-resistance (See other articles in these Proceedings.), which is now commercial in disk storage. Existing research programs already seek to exploit these advantages of TMR in recording-head sensors and non-volatile random-access memory, as detailed elsewhere in these Proceedings.

The second breakthrough, currently in the making, concerns EDME which is the main subject of this article. In essence, EDME replaces the Oersted-Maxwell magnetic field induced by electron-charge current with the s-d exchange field induced by electron-spin current as the agent for switching magnetization. Since an induced magnetic field is weak but spread over a large volume, whereas the exchange field is strong but concentrated within the small atomic cell, the exchange mechanism dominates at sufficiently small cross-sectional scales, circa $\leq 2$ nm thick $\times$ 200 nm wide. Just 3 years ago, L. Berger and the author independently predicted such excitations of one-dimensional spin-waves [4] and monodomain precession and reversal [5] by means of the exchange field created by steady spin-polarized electric current. Two years passed without any experimental tests of these predictions.

But at last, the past $1\frac{1}{2}$ years have seen several corroborative experiments. Reports appeared of observations [6,7] and an interpretation [8] of spin-wave instability in two dimensions excited through point contacts, and also of observations of domain reversal [7,9,10,11]. In metallic multilayers at low and ambient temperatures, the observed effects require circa 1 mA of steady electric current; in manganite particles at low temperature, less than 10 $\mu$A [10].

This report reviews the physics of EDME and presents some device concepts relying on EDME alone and in combination with TMR, which take advantage of its enormous switching efficiency.

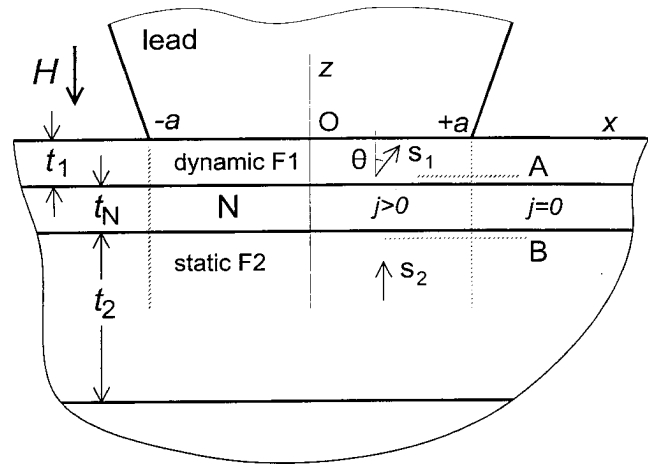

Fig. 1. Section of ideal contact region between an electric lead and the surface of a layered magnetic material [8]. Oz is the axis of rotational symmetry. F1 is the dynamic metallic magnetic layer and F2 is the static metallic magnetic layer. The layer marked N is a nonmagnetic metallic spacer. The current is assumed to flow within a cylinder of radius a.

**Theory [5,8].** Consider two infinitely extended parallel ferromagnetic layers F1 and F2 separated by a nonmagnetic metallic spacer N illustrated in Fig. 1. A nonmagnetic metallic lead, representing an experimental point contact, connects with the upper magnet F1 through a circular area of radius $a$. (This description also includes the different case of a cylindrical pillar like one at the top of Fig. 8, in which both current and magnetic layers are confined within the cylinder of radius $a$.) An effective magnetic field along the normal direction $-z$, generally combining external and anisotropy terms, saturates the magnetizations of both layers. The problem is to find the critical value of steady electric current $I$ flowing through the contact in the direction $+z$ which is needed for the dynamic micromagnetic excitation of magnet F1.

The exchange-reaction torque created by scattering of preferentially polarized electrons incident from a normal metal onto a ferromagnet is concentrated on those local ferromagnetic spins located within a distance equal to a few atomic layers of the interface [4,5]. The origin of this exchange torque is understood with the help of a brief digression to Fig. 2 which depicts graphically the local spin vector of the stationary-state wave-function for an electron passing rightward through an N/F interface located at position $\xi = 0$. (The coordinate axes $\xi\eta\zeta$ in Fig. 2 have no special relation to $xyz$ in Fig. 1.) Within the region N $(\xi < 0)$ its spin vector $s$ is a general constant. Upon passing into the ferromagnet $(\xi > 0)$ its azimuthal angle with respect to the moment $M$ of F is written

$$\varphi(\xi) = \varphi(0) + (k_{\xi+} - k_{\xi-})\xi, \quad (\xi > 0)$$

where $k_{\xi\pm}$ are the normal components of spin-up and spin-down Fermi vectors [5]. (For graphical simplicity, Fig. 2 shows $M \parallel \zeta$, but actually $M$ may have any direction.) For a given Fermi energy, $k_{\xi+}$ and $k_{\xi-}$ depend on the common conserved transverse component $k_\perp$ lying in the $\eta\zeta$ plane and differ in value because of s-d exchange. Since the electron wave vectors lie on all parts of the Fermi surface with diverse $k_\perp$, the averages of $s_\xi \propto \cos\varphi$ and $s_\eta \propto \sin\varphi$ approach 0 within an impact depth $d \approx 10$ or so atomic layers. It follows also that the momentum impulse given to F lies, on the average, within the original M-s plane. Thus, when electric current flows the reaction of this precession communicates to the magnet the net of the initial spin momentum components transverse to $M$ of *all* of the electrons entering the magnet F.

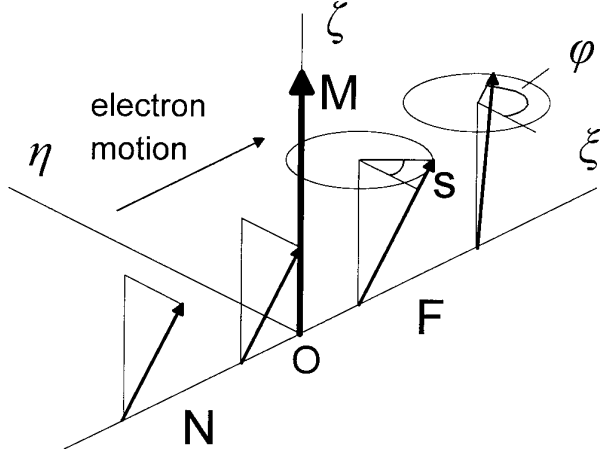

Fig. 2. Illustration of spin precession for an electron passing from a nonmagnetic metal $(\xi < 0)$ into a ferromagnetic metal $(\xi > 0)$.

Returning our attention to Fig. 1, let us assume that the scale of micromagnetic inhomogeneity treated in the continuum representation with the Landau-Lifshitz equations is greater than this impact depth $d$. Also let $s_1(t)$ and $s_2(t)$ be the instantaneous unit local 3d spin-moment vectors of magnets F1 and F2 evaluated at their interfaces with the spacer N. Then the above conclusions imply that current-induced exchange is mainly a surface effect and conservation of angular momentum requires the corresponding effective vectorial surface-torque densities $L_1$ and $L_2$ (with $L_n \cdot s_n = 0$) to satisfy the vector equation

$$L_1 + L_2 = K_2 s_2 - K_1 s_1 . \qquad (1)$$

Here $K_n \equiv \hbar(-j_{n\uparrow} + j_{n\downarrow})/2e$ is the upward flowing *spin-momentum* current (or spin current for brevity) density within magnet $n$, $-e$ is the electron charge and $j_{n\updownarrow}$ is the *electric* current density flowing in the up-or-down-spin (relative to the local $s_n$) channel as in the two-channel model of perpendicular magnetoresistance [12]. By convention, both charge and momentum *current* directions are reckoned positive along the $+z$ direction in Fig. 1. (The components $x$ and $y$ of electron movement are neglected.) Accordingly, the right-hand side of Eq. (1) represents the net rate of spin momentum flowing *into* the region enclosed by two horizontal geometric planes A and B (see Fig. 1) located inside the magnets at the distance $d$ from the F/N interfaces; the left side gives the consequent sum of macroscopic torques concentrated on the magnets at these interfaces.

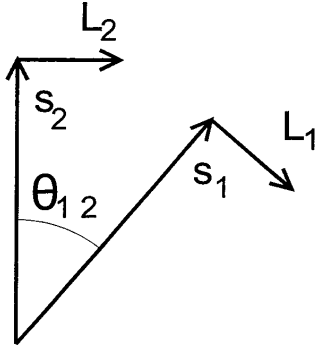

Fig. 3. Instantaneous common plane of local (3d) magnet spin moments $s_1$ and $s_2$ evaluated at adjoining ferromagnetic interfaces with a common paramagnetic spacer. Shown are the sign conventions for surface torques $L_1$ and $L_2$ acting on magnets F1 and F2 respectively (See Fig. 1 and Eqs. (2) and (3)). For magnets having equal polarizations $P$, WKB theory predicts $L_1 = L_2$.

The *coplanar* orientation of the axes of $\mathbf{L}_n(t)$ displayed in Fig. 3 are general in our macroscopic model. Their magnitudes are obtained by forming the scalar products of Eq. (1) alternatively with $s_1$ and $s_2$:

$$L_1 = (K_1 \cos\theta_{12} - K_2)/\sin\theta_{12},$$
$$L_2 = (K_2 \cos\theta_{12} - K_1)/\sin\theta_{12}, \quad (2)$$

where the sign convention for the scalars $L_n$ is indicated in Fig. 3.

Equations (1), or equivalently (2), are key to EDME for, given the spin currents $K_1$ and $K_2$, these equations predict the corresponding torques $\mathbf{L}_{1,2}$ included in the Landau-Lifshitz equations used below to treat *macroscopic* spin waves (as distinguished from thermal magnons) and domain switching. A detailed free-electron WKB treatment [5] of the spin currents in the special case where F1 and F2 have equal Fermi-level spin polarizations $P$ ($|P| \le 1$) and carry charge current density $j$, predicts the torques

$$\mathbf{L}_{1,2} = \hbar e^{-1} j g(\theta_{12}) \mathbf{s}_{1,2} \times (\mathbf{s}_1 \times \mathbf{s}_2), \quad (3)$$

with

$$g = \frac{\text{sgn } P}{-4 + \frac{(1+|P|)^3(3+\cos\theta_{12})}{4|P|^{3/2}}}. \quad (4)$$

(From tunneling studies one has $P = 0.40, 0.35, 0.23,$ and $0.14$ for Fe, Co, Ni, and Gd, respectively.)

Although the calculation [5] considered only $P > 0$, a simple consideration shows that only the sign of $g$ is affected by changing the sign of $P$. One may transform thus: Simultaneously reverse the sign of s-d exchange and the vectors $s_1$ and $s_2$. Now the two-component WKB wave functions derived in [5] are unchanged, and therefore the same torques $L_{1,2}$ are derived. Finally, rotation in the $s_1 - s_2$ plane through the angle $\pi$ restores $s_1$ and $s_2$ but gives $L_{1,2}$ opposite signs from before, showing that the factor sgn $P$ in Eq. (4) is correct.

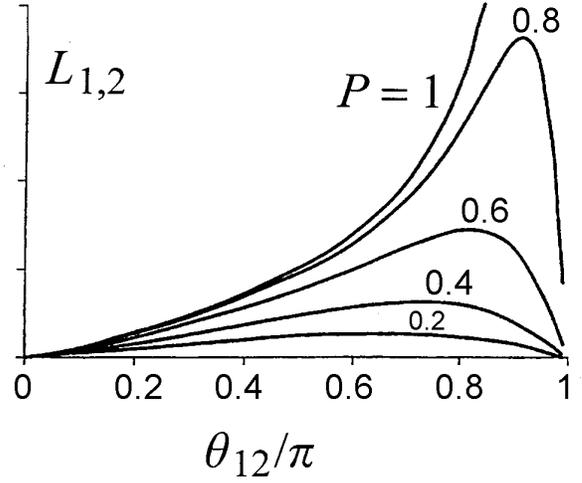

Fig. 4. Torques (in arbitrary units) created by current-induced exchange versus angle included between two magnets. WKB theory with both magnets having spin polarization $P$, according to Eqs. (3) and (4) [5].

Note the curious equality $L_1 = L_2$, both in magnitude and sign (with the convention of Fig. 3) for the WKB model. The current impels the vectors $s_1$ and $s_2$ to rotate *in the same sense* within their instantaneous common plane. The amplitude of these torques is plotted in Fig. 4. (Our illustration of the case $P > 0$ does not accord with computed band structures but rather, because of the elementary nature of the free-electron picture, with experimental data for most ferromangets showing that majority-spin electrons carry more current than minority-spin electrons.) For half-metallic ferromagnets ($P = 1$) the torques reduce to

$$\mathbf{L}_{1,2} = \hbar j \mathbf{s}_{1,2} \times (\mathbf{s}_1 \times \mathbf{s}_2)/2e(1 + \mathbf{s}_1 \cdot \mathbf{s}_2). \quad (5)$$

The WKB-model expressions (3-5) have the virtue of one-parameter ($P$) simplicity. A quantitative interpretation of experiments might require more detailed calculation of the four $j_{n\updownarrow}$ in some extension of the two-channel resistive network model [12] to the present case of non-collinear moments. This should affect the functional dependence of $g(\theta_{12})$ and perhaps make it depend on $n = 1, 2$ without otherwise changing the form of Eq. (3).

Note that the relations (1) and (2) are incompatible with any naive picture of electrons moving only in the direction $-j$, which falsely suggests that only one direction of current can excite a given one of the two magnets. The properties predicted by these relations only make sense when one considers that flow of current represents a usually small directed drift superposed on a tempest of multiply scattered electron motions in all directions. In the WKB model [5], electron flow in the direction F1 → F2 creates torque on F1 because electron waves partially reflect at N/F2 and N/F1 interfaces. The greater degree of reflection near $\theta_{12} = \pi$, associated with greater (magneto-)resistance, than near $\theta_{12} = 0$ explains the inequality

$$|dL_{1,2}/d\theta_{12}|_{\theta_{12}=\pi} > |dL_{1,2}/d\theta_{12}|_{\theta_{12}=0}$$

apparent in Fig. 4.

Calculation of the dynamic effect of exchange due to spin-polarized current requires inclusion of the surface torques given by Eqs. (2), (3) or (5) in the conventional Landau-Lifshitz equation.

**Threshold current for instability of magnetostatic equilibrium.** With reference to Fig. 1, now assume F2 is static because of great thickness and regard F1 as two-dimensional with $d \approx t_1$ small. For a two-dimensional spin wave in F1 with circular precession expressed by $s_{1x} + is_{1y} = e^{i\omega t} u(x, y)$, the *linearized* Landau-Lifshitz equation, including the current-driven torque of Eqs. (3) and (4), reduces to

$$\left[\omega - \gamma H_{\text{eff}} + \frac{2\gamma A}{M_{s1}}(\partial_x^2 + \partial_y^2) - ia_G\omega + i\frac{\hbar\gamma g(0)}{eM_{s1}t_1}j(x,y)\right]u = 0, \tag{6}$$

$$\text{with} \quad H_{\text{eff}} = H - 4\pi M_{s1} + H_{\text{ex}} + H_u. \tag{7}$$

Here the units are cgs emu with gyromagnetic ratio $\gamma$, external field $H$, spontaneous magnetization $M_{s1}$, effective field $H_{\text{ex}}$ of a possible exchange coupling to magnet F2, uniaxial material anisotropy field $H_u$, coefficient $A$ of exchange stiffness energy density $A|\nabla\theta|^2$, and Gilbert viscous-damping coefficient $a_G$.

The successive terms in this equation attribute the circular precession of moment $s_1(x, y, t)$ at frequency $\omega$ to effective fields of in-phase type (real) due to $H_{\text{eff}}$ and exchange stiffness, and of quadrature type (imaginary) due to viscous dissipation and above-derived current-driven exchange torque. Whereas viscous dissipation can only dampen the motion ($a_G > 0$), the effect of spin injection depends on the sign of the electric current. If the electric current flows upward in Fig. 1, then one has $j > 0$ in Eq. (6) and damping is opposed so that excitation of spin waves becomes conceivable. The steady current creates an effective rf field which is always collinear, in effect "at resonance", with the precession $d\boldsymbol{M}_1/dt$ for any frequency. As implied by Fig. 3, this "resonance" persists generally even as the precession frequency varies with amplitude in a nonlinear motion.

To model point contacts, one considers $j$ to be constant within a cylinder of radius $a$ and to vanish outside of this cylinder. An initial uniform static state with $I \equiv \pi a^2 j = 0$ constitutes a stable dynamic fixed point. Raising $I$ above a critical value makes this point unstable. A Bessel-function solution of Eq. (6) describing the symmetric radiation of spin waves predicts the critical-current expression

$$I_{\text{crit}} = \frac{et_1}{2\hbar g(0)}(23.4A + 6.31a^2 a_G M_{s1} H_{\text{eff}}) \tag{8}$$

for excitation out of the uniform $s_1 = s_2$ state [8]. (This expression is conveniently regarded in mixed units with $I_{\text{crit}}$ and $e$ in SI and the other quantities in cgs emu.) For excitation out of the possibly stable state $s_1 = -s_2$ (if, for example, $H = H_{\text{ex}} = 0$ and $H_u > 4\pi M_s$) the sign of the current reverses and $g(\pi)$ replaces $g(0)$.

The two terms in Eq. (8) represent physically distinct contributions to the threshold. The *initial* ($H_{eff} = 0$) critical current $I_{crit\,0} = 23.4eAt_1/2\hbar g(0)$ given by the first term is due to spin-wave radiation of energy away from the central excited region. Its independence of excitation area $\pi a^2$ reflects the fact that the excited wavelength is determined by the contact radius independently of field. Since the group velocity for spin waves varies inversely with this wavelength, the effectiveness of spin waves to radiate energy decreases with increasing $a$. It happens that, at a *given total current*, the resulting decrease in effectiveness of energy transport exactly compensates for the corresponding decrease of injected torque density, leaving the radiative contribution to the critical current unchanged.

The second term in Eq. (8) is due to local viscous damping of the spin precession. This field-dependent contribution to *critical current density* $j$ does not essentially depend on whether the excited magnet is *bounded* within an area equal to that of the current, or unbounded. Indeed, it follows almost exactly from equating to 0 the sum of the last two terms of Eq. (6) and substituting $\gamma H_{eff}$ for $\omega$. Evaluation of $\alpha_G$ in Eq. (8) needs care because of a predicted additional interfacial contribution which is generally competitive with that valid in ordinary ferromagnetic resonance [4]. Expressions along the line of this viscous-damping term alone are currently used to interpret experimental switching of manganite particles [10] and Co films in lithographed pillars fashioned by lift-off (See below) [11].

**Monodomain switching.** To treat what happens when the current exceeds the derived threshold [Eq. (8)] will require solution of the Landau-Lifshitz equations for general $\theta$ without linearization. For simplicity, consider a uniformly magnetized monodomain having uniaxial effective anisotropy field $H_u = 2K_u/M_{s1}$, where $K_u \sin^2\theta$ is the total energy per unit volume, including material and shape terms [5]. The free motion of the monodomain is a circular precession about the easy axis with constant $\theta$ and circular frequency $\omega = \gamma H_u \cos\theta$. In the presence of small damping and exchange torques, the time-dependence of the cone angle satisfies

$$d\theta/dt = \dot\theta_{damp} + \dot\theta_{exch} \qquad \text{with}$$

$$\dot\theta_{damp} = -\frac{1}{2}\gamma aH_u \cos 2\theta \qquad \text{and}$$
$$\dot\theta_{exch} = \frac{\gamma\hbar jg(\theta)}{eM_{s1}t_1}\sin\theta. \qquad (9)$$

The latter three functions are plotted in Fig. 5 for three values of current $I = -1, -2, -4$. (Units for all physical quantities are arbitrary.) Obvious conditions for the stability of any cone angle are $d\theta/dt = 0$ (equilibrium) and $d[d\theta/dt]/d\theta < 0$ (stability).

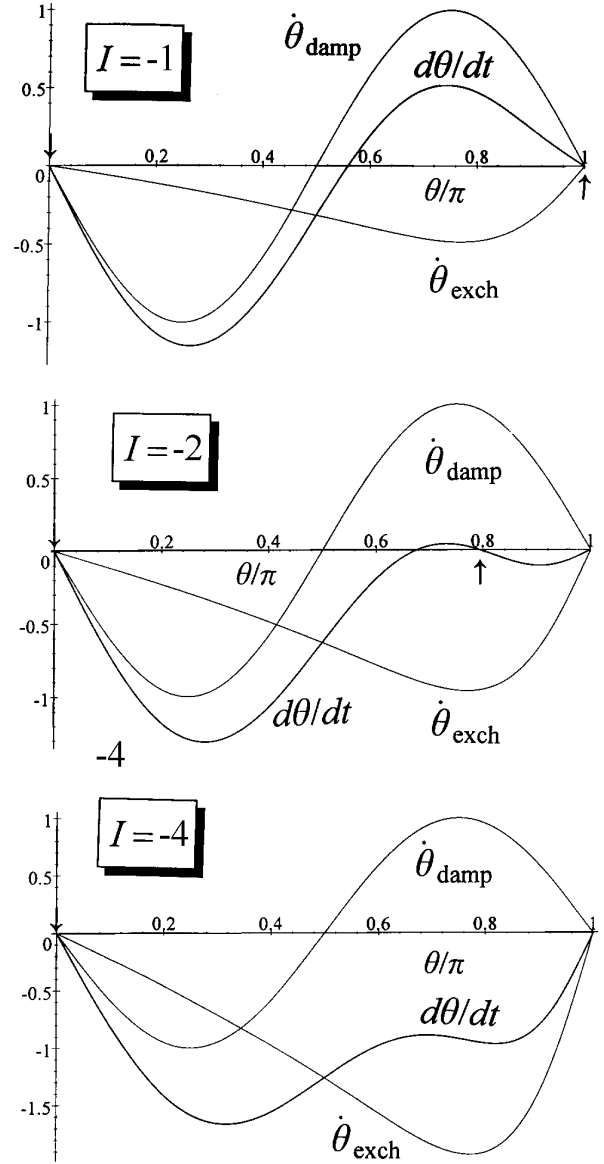

Fig. 5. Instantaneous angular velocity versus angle $\theta$ of the precession cone for a uniaxial monodomain magnet F1 subject to viscous damping and three values of dimensionless current $I$, according to Eqs. (9). Points of stable dynamic equilibrium are indicated by ↑ or ↓.

For $I = -1$ the remanent states $\theta = 0, \pi$ satisfy both conditions, but the intermediate equilibrium point $\theta = 0.56\pi$ is not stable. Therefore the current value $I = -1$ does not excite either of the two remanent states. In the time domain, consider a small initial fluctuation (e.g. thermal) $\theta = 0.95\pi$ from one remanent state. Then, integration of Eqs. (9) shows that the current $I = -1$ permits the moment to relax

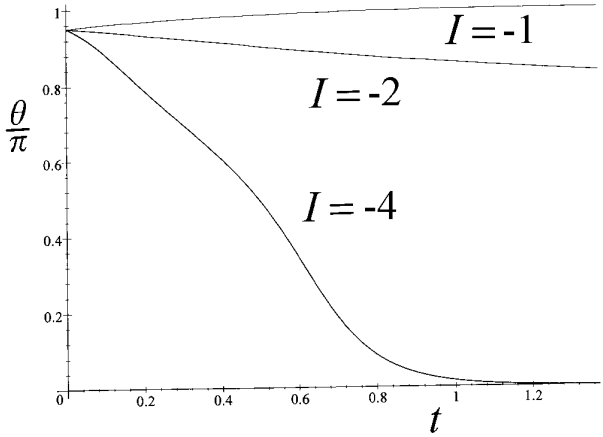

Fig. 6. Dependence of precession-cone angle $\theta$ on time $t$ computed from Eqs. (9), in arbitrary units. The initial state is $\theta = 0.95\pi$. The current $I = -1$ causes no switch, $I = -2$ causes a partial switch to the precessing state $\theta = 0.79\pi$, and $I = -4$ causes a full switch to $\theta = 0$.

exponentially to the nearby remanent state as illustrated by the top curve in Fig. 6.

For $I = -2$, only two, ($\theta = 0, 0.79\pi$) of the four equilibrium states are stable. Therefore this value of current drives the moment out of the neighborhood of $\theta = \pi$ toward the first stable equilibrium $\theta = 0.79\pi$. (See Fig. 6.) After it relaxes to this point, the moment continues to precesses steadily at circular frequency $\omega = \gamma H_u \cos 0.79\pi$ as long as the constant current $I = -2$ is maintained. If the current is subsequently turned off, then the moment falls to the nearer remanent state, $\theta = \pi$ in this case. This example illustrates the fact that the criterion $d[d\theta/dt]/d\theta = 0$ for instability does not necessarily imply a full moment reversal.

For $I = -4$, only one state has a stable equilibrium, so that a complete reverse switch from $\theta = \pi^-$ to $\theta = 0$ occurs. (See Fig. 6.) Note that the current speeds up the relaxation to the final state.

Clearly, positive $I$ of sufficient magnitude will switch in the forward direction from $\theta = 0^+$ to $\theta = \pi$. In this case ($I > 0$) it happens that a steady precessing state does not exist and the threshold current for instability of $\theta = 0$ does also cause a full switch. Thus for $H_u > 0$, the possibility of a steady precessing state depends on the value of $P$ and the sign of $I$. However, for $H_u < 0$ there exists a range including both signs of $I$ supporting a steady precession [5].

**Experiment.** Investigation of EDME is made difficult by the need for at least one small lithographic width defining the film-normal current flow. If this width greatly exceeds about 200 nm, then the magnetic field induced by the current disturbs the otherwise possibly uniform micromagnetic state, thus potentially obscuring the desired exchange-induced effect.

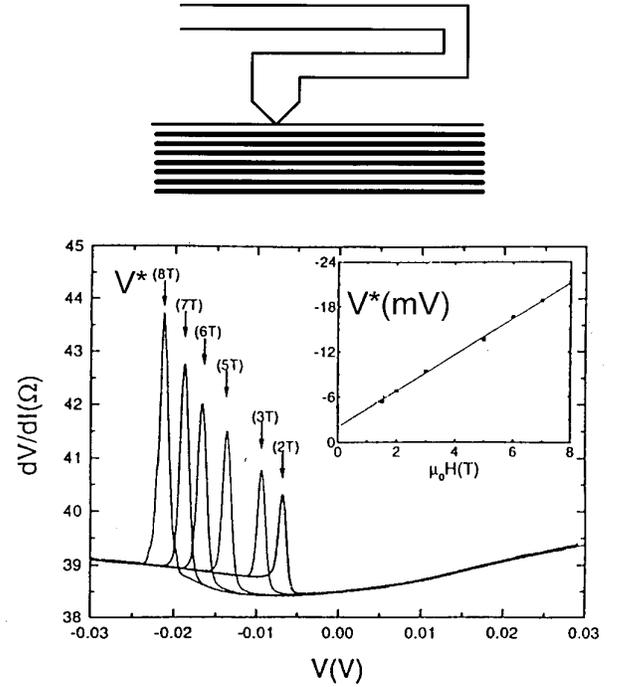

Fig. 7. Mechanical point contact to multilayer (top) and experimental current-induced V/I anomaly (bottom) observed for different plane-perpendicular fields. Inset shows field-dependence of peak position in dc voltage $V^*$ (and therefore current) [6].

The I/V anomaly plotted in Fig. 7 is the first published evidence of exchange-driven magnetic excitation, by M. Tsoi et al [6]. A mechanical point contact supplies the current which flows through a multilayer of composition (Co 1.5nm/Cu 2nm)$_N$ with $N = 20 - 50$ at a temperature of 4.2 K. (See Fig. 7, top.) In some cases the position of the observed voltage anomaly $V^*$, which signals a sudden change in CPP-GMR (current perpendicular to the plane giant magnetoresistance [13]), shows the linear field dependence of $I_{crit}$ (See inset in Fig. 7) expected from the above (subsequently derived) 2-layer theory [See Eq. (8)]. Thus, in spite of the large number of experimental layers, the data sometimes support both spin-wave radiation and damping contributions to the threshold of the two-unequal-layer model.

A similar experiment by E. B. Myers et al at Cornell University [7] used electron-beam lithography and reactive ion etching to produce the point contact. Just 2 cobalt layers, one 100 nm thick and the other between 0 and 10 nm, in accord with the above theoretical model, were sputtered. The contact diameter varied between 5 and 10 nm. Now, a more consistent linear dependence of threshold current on external field at 4.2K occurred, but with the $H_{eff} = 0$ intercept still 2 to 4 times lower than predicted (neglecting $H_{ex}$ and $H_u$ in Eq. (7)). Also, the value of $\alpha_G$ consistent with the slope is considerably larger than that known independently from ferromagnetic resonance of MBE cobalt films. In these high-field experiments, a strong evidence of the general correctness of the exchange-driven mechanism is that the anomaly occurs for only one sign of current (See Fig. 7), as predicted. Any effect caused by induced magnetic fields would be symmetric in the sign of the current.

Still more recent work at Cornell [11] does much to resolve puzzling aspects of the physics and advance greatly the device potential of EDME. The top of Fig. 8 illustrates a $130 \pm 30$ nm diameter pillar, fabricated with electron-beam lithography and lift-off, which contains cobalt films of thickness 2.5 nm (F1 = Co1) and 10 nm (F2 = Co2) separated by 6 nm of Cu. The plot in the center shows differential resistance versus current in the presence of 1.2 and 1.6 kOe external fields applied parallel to the film planes. (The 1.2 kOe plot is displaced upward for clarity.) The temperature is ambient. The quasi-rectangular hysteresis loops demonstrate convincingly monodomain switching in Co 1, as supported by the magnetic hysteresis of simple resistance shown in the bottom plot. (Evidence of one unexplained two-step transition appears in both types of hysteresis.) The two plots show nearly the same total resistance change 72 m$\Omega$.

Surely the I/V relation observed in the center of Fig. 8 cannot be attributed to self-induced magnetic field. For, any magnetoresistive effect of in-plane magnetic field induced by current flowing perpendicular to the plane must be broadly symmetric with respect to $I \rightarrow -I$. An experimental example exhibiting this symmetry is the very recent, otherwise similar observation of inductive switching of magnetic vortices circulating about the center of a many-layered circular multilayer [14].

To explain the data of Fig. 8 with current-driven exchange fields, consider that the confinement of a magnet to a small pillar provides no means of spin-wave radiation. Thus the first term of Eq. (8) is omitted. Since the external field needed to orient the moment of F2 is parallel to the film plane, the precession in F1 is highly elliptic rather than circular as assumed in deriving Eq. (6). For an F1 volume $V_1$ in this case, a solution of the Landau-Lifshitz equation including the torques (3) and (4) predicts the two thresholds

$$I_{\text{crit}} = a_G e M V_1 [H_{\text{eff}}\binom{0}{\pi} \pm 2\pi M_{s1}]/\hbar g\binom{0}{\pi} \quad (10)$$

now replacing Eq. (8) [11]. The large new term $\pm 2\pi M_{s1} \simeq \pm 8$ kOe is introduced into Eq. (10) by the drastic elliptic distortion of the magnetic precession caused by the great z-axis demagnetization. Despite this fundamentally unnecessary hindrance, the switching thresholds $\approx 5 \pm 4$ mA observed in the 1.6 kOe loop of Fig. 8 are modest. (Additional measurements use the shape anisotropy produced by 120 nm x 60 nm lithographed dimensions instead of external field to orient the moments statically [15]. They unambiguously reveal sharply defined switching thresholds of 2.2 and −1.8 mA.)

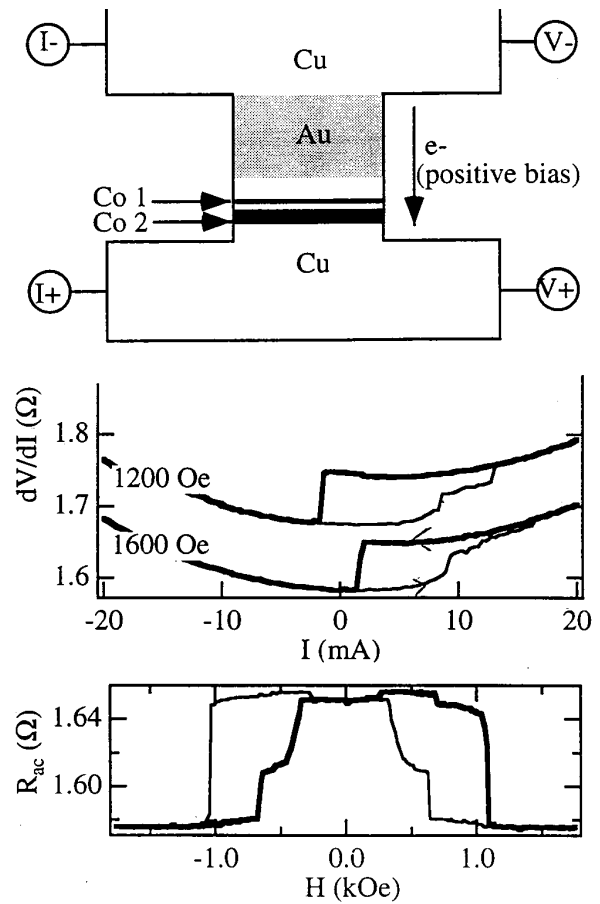

Fig. 8. Pillar structure (top) and experimental current-induced monodomain switching hysteresis (center) of film Co1. The field is parallel to the layer plane and detection is by differential resistance [11]. The bottom plot shows that, for $I = 0$, Co1 switches in the field ranges $\pm(0.3$ to $0.6)$ kOe.

At fields $H < 1$ kOe, where a vortex may occur in Co 2, the experimental I/V behavior [11] is unexplained. At higher fields in the range 1.6 kOe $< H <$ 27 kOe the I/V behavior indicates incomplete switching. Only one rather than two critical thresholds exists at these higher fields. The remarkably small, nearly constant slope of its field dependence $dI_{\text{crit}}/dH = 0.29$ mA/kOe is fairly consistent with Eq. (10) and the loops shown in Fig. 8. A fit of Eq. (10) to this slope uses the adjusted damping $a_G = 0.005$ and independently known values of the other magnetic parameters for cobalt [11]. A ferromagnetic resonance value for MBE cobalt is $a_G = 0.007$, supporting approximate consistency. This better agreement with theory of pillar data as compared to point-contact data is attributable to better definition of the current distribution.

**Device concepts.** This new fact of experimental confirmation of EDME and its high potential for switching-current economy stimulate consideration of device concepts. Figure 9 illustrates a magnetic write stylus, incorporating the permanent-magnet spin polarizer F1 (in notation differing from that of Fig. 1) to be used in place of an inductive write head in magnetic recording [16]. The slanted layer structure S is a soft magnetic shield to decrease disturbances caused by the stray induced magnetic field.

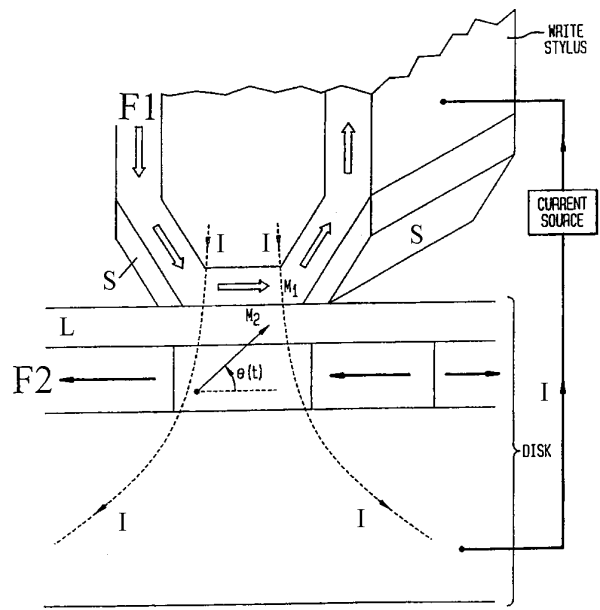

Fig. 9. Electric write-stylus device concept for magnetic recording [16]. F1 is the polarizing permanent magnet. F2 is the switchable storage medium. S is a magnetic shield which diminishes the disturbing effect of current-induced magnetic fields.

The write stylus of Fig. 9 naively features a conducting lubricant L to enable current flow between F1 and the storage cells F2 on disk. Commercial use of EDME in writing must await a major invention, perhaps one utilizing a microscopic electron gun or a spark discharge, to allow the write probe to pass 1 ns, 10 mA current pulses to the medium without mechanical contact. Such an invention could enable EDME to overcome existing limitations on write rate due to inefficiency of today's advanced inductive heads requiring 1 ampere-turn. In addition, the present limit on storage density due to superparamagnetism may be raised by using storage particles whose anisotropy field exceeds the limitation presently imposed by the gap-fringe field of a saturated write-head core.

The TMR and EDME phenomena complement each other. Combining them into one 100 nm-sized integrated structure should provide an active high-speed 3-terminal device [16]. The current in the low-resistance input loop of the rotationally symmetric device of Fig. 10 is spin-polarized by the permanent magnet F1 (in notation differing from that of Fig. 1). Much of the spin current emanating from F1 flows to the soft magnet F2 and excites it through quantum-mechanical exchange between conduction and local d electrons. The magnetic orientation of F2 in turn controls a large resistance in the output loop by virtue of TMR provided by permanent magnet F3, thereby providing an output voltage signal with gain. (The theoretical expectation that electron back-scattering will excite F2, *without net charge current actually passing through F2 and the barrier,* needs experimental verification.)

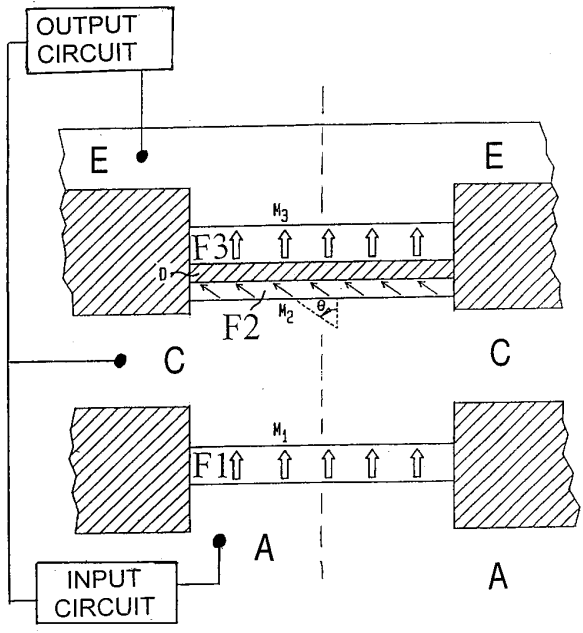

Fig. 10. Section of a conceptual generic rotationally symmetric three-terminal active magnetoelectronic device [16]. It incorporates tunneling magnetoresistance together with exchange-driven magnetic excitation. F1, F2, and F3 are ferromagnetic metallic films. D is a nonmagnetic tunneling barrier. Other shaded regions are of insulating material. A, C, and E indicate contacts to three copper layers.

The function of the Fig. 10 device depends on the combination of anisotropy directions and magnitudes provided in the 3 magnetic layers. Figure 11 indicates schematically three possible device functions:

(A) **latch.** In a non-volatile latch, suitable for memory, all three magnets have the same common easy axis (vertical in the example of the figure) for the net of material and shape anisotropy fields. The input current pulse $I(t)$, flowing as in Fig. 10, switches F2, rather like at the center of Fig. 8 but with conditionally smaller threshold currents whose signs depend on the direction of switch. It thereby changes the resistance of the tunnel barrier in the output circuit, which provides sensing of the latched condition.

Consider the 0.29 mA of *transmitted* current per kilo-Oersted of external orienting *magnetic* field needed in the above-discussed Cornell experiment to excite a 130 nm diameter film away from magnetic equilibrium. Assuming comparable current-to-torque conversion efficiency in the *back-scattering* mode needed by the device in Fig. 10, an equal current per kilo-Oersted of net effective plane-perpendicular *uniaxial* anisotropy field coefficient $H'_{u2} = H_{u2} - 4\pi M_{s2}$ would switch F2 from parallel to antiparallel as described above. (Quasistatic switching of such an uniaxial element using the *magnetic* field induced by a 130 nm wide wire would require almost 100 times as much current for the same $H'_{u2}$. Note that the huge $\pm 2\pi M_{s1} = \pm 8$ kOe shape-anisotropy term in Eq. (10) for the threshold current, which contributes greatly to the in-plane switching threshold currents in Fig. 8, is absent in the uniaxial case.) Thus, a coefficient $H_{eff} = H'_{u2} = $ 1-3 kOe for switching along the perpendicular direction would provide a smaller write-current threshold ($\approx 0.3$-$0.9$ mA) according to the first term of Eq. (10).

This large magnitude of $H'_{u2}$ compared to the switching field ($\approx 50$ Oe) typical of in-plane magnetized memory elements implies increased memory stability against troublesome micromagnetic and thermal-activation effects. Such effects are caused by vortices and pinning of edge solitons due to stray fields and roughness at edges of devices using existing in-plane magnetic-field write schemes. Indeed, the relative weakening of self-demagnetizing fields at the edges of the proposed device tends to *increase* the stability and remanence ratio of the information states.

Rigorous theoretical analysis of the initial phase of exchange-driven switching for films is not available but, once the precession cone angle exceeds some small threshold, the reversal clearly proceeds, at the 100 nm lithographic scale, with an angular rate as high as 1GHz per milli-Ampere of drive in excess of threshold [5,16]. Unlike the case of Maxwell fields, the scaled current density does not increase with increasing integrated-device density. Note that, considered as a logic gate rather than a memory element, the utility of such a latch would differ from that of silicon devices because of its reliance on current rather than voltage for input.

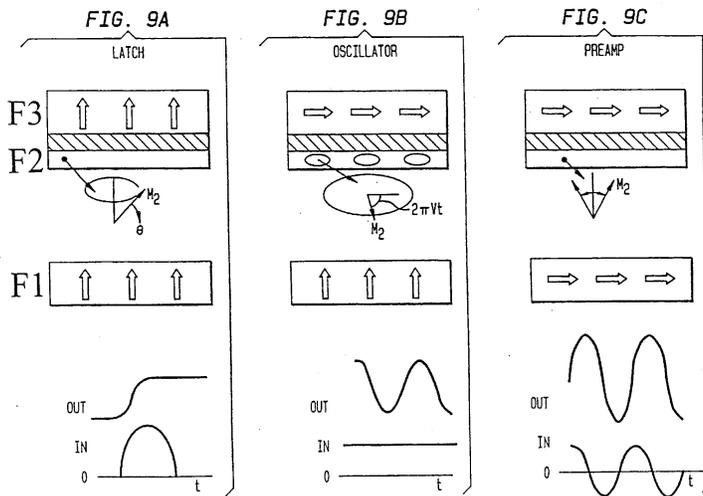

Fig. 11. Schematic depictions of three device functions depending on anisotropies of the three magnets F1, F2, and F3 shown in Fig. 8 [16].

B) **oscillator.** To obtain the oscillator function indicated in Fig. 11B, let the uniaxial anisotropy of F2 be easy-plane ($H'_{u2} < 0$). A dc input current exceeding a presently undetermined threshold causes a steady circular precession of $\mathbf{M}_2$ [5]. This provides a sinusoidal output voltage which is driven and frequency-tuned by the input current. Such an oscillator should have a broad frequency band of many giga-Herz [16].

C) **amplifier.** To produce an analog amplifier indicated in Fig. 11C, let the common easy axes of anisotropy of F1 and F3 be orthogonal to the easy axis of F2. A continuous input current produces response in F2, thus producing a phase-shifted output voltage [16].

For the future, a theoretical treatment of exchange-field switching by backscattering is still needed. The new material challenges posed by this new physics of TMR and EDME include the development of (1) tunneling barriers with lower (resistance × area) and (2) very thin magnetic films having both small losses and tailored perpendicular magnetic anisotropy.

The author is happy to thank D. Ralph and J. Katine for access to their research results prior to publication, and for permission to reproduce figures from their preprint. Grateful thanks are also due to J. Sun and R. Koch for discussions and to R. Matick and N. Garcia for comments on a previous draft of this manuscript.

## References

1. M. Julliere, Phys. Lett. **54A**, 225 (1975).

2. J. C. Slonczewski, IBM Technical Disclosure Bulletin **19,** 2328; 2330 (1976).

3. T. Miyazaki and N. Tezuka, J. Magn. Magn. Mater. **139** L231 (1995): J. S. Moodera, L. R. Kinder, T. M. Wong, and R. Meservey, Phys. Rev. Lett. **74,** 3273 (1995) See also these Proceedings.

4. L. Berger, Phys. Rev. B 54, 9353 (1996); J. Appl. Phys. 81, 4880 (1997); IEEE Trans. Magn. 34, 3837 (1998).

5. J. C. Slonczewski, J. Magn. Magn. Mater. 159, L1 (1996).

6. M. V. Tsoi, A. G. M. Jansen, J. Bass, W.-C. Chiang, M. Seck, V. Tsoi, and P. Wyder, Phys. Rev. Lett. 80 (1998) 4281.

7. E. B. Myers, D. C. Ralph, J. A. Katine, R. N. Louie, and R. A. Buhrman, Science 283, 867 (6 Aug., 1999).

8. J. C. Slonczewski, J. Magn. Magn. Mater. 195, L261 (1999).

9. J.-E. Wegrowe, D. Kelly, Y. Jaccard, Ph. Guittienne, and J.-Ph. Ansermet, Europhys. Lett. 45, 626 (1999). See also these Proceedings.

10. J. Sun, J. Magn. Magn. Mater. 202, 157 (June 1999).

11. J. A. Katine, F. J. Albert, R. A. Buhrman, E. B. Myers, and D. C. Ralph, submitted for publication.

12. T. Valet and A. Fert, Phys. Rev. B 48, 7099 (1993).

13. See J. Bass, W. P. Pratt Jr., P. A. Schroeder, Comments on Condensed Matter Phys. 18, 223 (1998).

14. K. Bussmann, G. A. Prinz, S.-F. Chang, and D. Wang, Appl. Phys. Lett. 75, (18 October, 1999) 2476.

15. J. Katine, private communication.

16. J. C. Slonczewski, U. S. Patent # 5,695,864 (1997).